\documentstyle[12pt]{article}

\begin{document}

\noindent{\bf QUANTUM OSCILLATOR WITH KRONIG-PENNEY}\\
\noindent{\bf EXCITATION IN DIFFERENT REGIMES OF DAMPING}\\

\indent Olga Man'ko\\

\indent Lebedev Physical Institute\\
\indent Leninsky prospekt 53, Moscow 117924, Russia\\
\indent fax: (095) 938 22 51, e-mail: manko@sci.fian.msk.su\\

\noindent{\bf INTRODUCTION}\\

The behaviour of an oscillator may be controlled by the frequency
time--dependence. For example, one can kick the oscillator frequency by
short pulses and this kicking produces an excitation of the parametric
oscillator. The amplitude of the oscillator vibrations and its energy may
increase due to the external influence expressed as the frequency
time--dependence. Also the statistical properties of the oscillator state
may be changed due to the action of external forces.
The aim of the talk is to discuss the exact solution of the time--dependent
Schr\"{o}dinger equation for a damped quantum oscillator subject to a
periodical frequency delta--kicks describing squeezed states which are
expressed in terms of Chebyshev polynomials. The cases of strong and weak
damping are investigated in the frame of Caldirola--Kanai model
\cite{[1]}, \cite{[2]}.

The problem of quantum oscillator with a time--dependent frequency was
solved in Refs. \cite{[3]}--\cite{[11]}. It was shown that the wave
function and, consequently, all physical characteristics of the
oscillator can be expressed in terms of the solution of the classical
equation of motion
\begin{equation}\label{eq.1}
\ddot \varepsilon (t) + 2\gamma \dot \varepsilon (t) + \omega^{2} (t)
\varepsilon (t) =0, \end{equation}
\noindent with initial conditions
\begin{eqnarray}\label{eq.2}
\varepsilon(0)&=&1,
\nonumber\\
\dot \varepsilon (0)&=&\dot \imath \Omega (0).
\end{eqnarray}
\noindent where $~\Omega (0)=~\Omega $ will be defined below.
The remaining problem is to find explicit expression for the function
$~\varepsilon .$\\

\noindent{\bf DIFFERENT REGIMES OF DAMPING}\\

Here we consider the case of a periodically kicked
oscillator, where the frequency depends on time as follows
$$\omega ^2 (t) = \omega^2_0 - 2\kappa \sum _{k=1}^{N-1} \delta (t -
 k\tau), $$
\noindent where $~\omega _0$ is constant part of frequency, $~\delta $
is Dirac delta--function, $~\gamma $ is the damping coefficient, and
$~\kappa $ is the force of delta--kicks. We consider the damping in the
frame of Caldirola--Kanai model, and take into account three cases:\\
(i) undamped case ($~\gamma =0$);\\
(ii) the case of weak damping ($~\omega _{0} > ~\gamma $);\\
(iii) the case of strong damping ($~\omega _{0} < ~\gamma $).\\
\noindent The undamped case was considered in \cite{[10]}; following
\cite{[10]} we have the equation for function $~\varepsilon (t)$
\begin{equation}\label{eq.3}
\ddot\varepsilon (t) + 2\gamma \dot\varepsilon (t)
+ \omega _0^2 \varepsilon (t)
-2 \kappa \sum_{k=1}^{N-1} \delta ( t - \kappa \tau ) = 0.
\end{equation}
\noindent It is obvious, due to substitutions $~t$ by $~x$,
{}~$\varepsilon $ by $~\psi $, and $~\omega _0^2 / 2$ by $~E$, that if the
damping is absent this equation coincides with the equation for the wave
function of a quantum particle of unit mass in a Kronig--Penney potential
(the sequence of $~\delta $--potentials). For every interval of time
$~(k-1)\tau <~t<~k\tau $ the solution for the classical equation of
motion is given by
\begin{equation}\label{eq.4}
\varepsilon _k (t) = A_k e^{\mu _{1} t} + B_k e^{\mu _{2} t},
{}~~~k=0,1,\ldots ,N,
\end {equation}
\noindent $~\mu _{1}$ and $~\mu _{2}$ are complex numbers. Due to
continuity conditions we have
\begin{eqnarray}\label{eq.5}
\varepsilon _{k-1} (k\tau) &=& \varepsilon _k (k\tau),
\nonumber\\
\dot\varepsilon _k (k\tau) -\dot\varepsilon _{k-1} (k\tau) &=& 2\kappa
\varepsilon _{k-1} (k\tau).
\end{eqnarray}
\noindent Formulae (\ref{eq.5}) are obtained by integrating
Eq. (\ref{eq.3}) over the infinitely small time interval
$~n\tau - 0 < ~t < ~n\tau + 0$.
The coefficients $~A_k$ and $~B_k$ must satisfy the relations which can
be expressed in the matrix form
\begin{equation}\label{eq.6}
\pmatrix{ A_k \cr B_k}
=\pmatrix{1-\frac{2\kappa }{D} &-\frac{2\kappa }{D}e^{D\tau k} \cr
\frac{2\kappa }{D}e^{-D\tau k} & 1+\frac{2\kappa }{D}} \pmatrix{
A_{k-1} \cr B_{k-1}},  \end{equation}
\noindent where $~D=~\mu _{2} - \mu_{1}$. After the sequence of
$~\delta $--kicks the coefficients $~A_n$, $~B_n$ are connected with
the initial ones $~A_0$, $~B_0$ through the equation
\begin{equation}\label{eq.7}
\pmatrix{ A_{n}\cr B_{n} } = S^{(n)} \pmatrix{ A_{0}\cr B_{0} },
{}~~~~~~~S^{(n)}=T^{-(N-1)}(M T)^{n},\end{equation}
\noindent with matrices $~T$ and $~M$ given by
$$T=\pmatrix{e^{-D\tau /2} & 0 \cr 0& e^{D\tau /2}}, ~~~
M=\pmatrix{1-\frac{2\kappa }{D} &
-\frac{2\kappa }{D} \cr \frac{2\kappa }{D} &
1+\frac{2\kappa }{D}}.$$
\noindent Thus the elements of the matrix $~S^{(n)}$ are of the form
\begin{eqnarray}
S_{11}^{(n)}& =&
(1-\frac {2\kappa }{D})U_{n-1}(\chi /2)e^{D\tau (n-2)/2} -
U_{n-2}(\chi /2)e^{D\tau (n-1)/2},
\nonumber\\
S_{12}^{(n)}&=&-\frac{2\kappa }{D}
U_{n-1}(\chi /2)e^{D n\tau /2},
\nonumber\\
S_{21}^{(n)}&=& \frac{2\kappa }{D}U_{n-1}(\chi /2)e^{-D n\tau /2},
\nonumber\\
S_{22}^{(n)}& =&
(1+\frac {2\kappa }{D}U_{n-1}(\chi /2)e^{-D(n-2)\tau /2} -
U_{n-2}(\chi /2)e^{-D(n-1)\tau /2}.
\end{eqnarray}
\noindent where $~U_{n-1}$, $~U_{n-2}$ are Chebyshev polynomials of the
second kind defined by the expression:
$$ U_n (\cos \varphi ) =
\frac{\sin(n+1)\varphi }{\sin \varphi };$$
\noindent with argument $~\chi /2=\frac {1}{2}\mbox {Tr}~MT$.

If at the initial moment of time the quantum oscillator was in a
coherent state the parametric excitation will transform it into a
squeezed correlated state with coordinate variances
$~\sigma _x(t)=~\frac{\hbar}{2m\Omega }~\mid\varepsilon \mid^2,$
and squeezing coefficient
$~K=~\frac {\sigma _x(t)}{\sigma _x(0)}=\mid \varepsilon \mid^{2}.$
Thus after the sequence of $~\delta $--kicks one has
\begin{eqnarray}\label{eq.9}
\sigma _x (t) &=& \mid A_{n}\mid^{2} \exp (\mu _1 + \mu _1^{\ast})t~
+ \mid B_{n}\mid^2 \exp (\mu _2+\mu ^\ast_2)t\nonumber\\
&+&B_{n} A_{n}^{\ast}\exp (\mu _2+\mu_1^{\ast})t
+ A_{n} B_{n}^{\ast}\exp (\mu _1+\mu _2^{\ast})t.
\end{eqnarray}
\noindent In the case of zero damping ( $~\gamma =0$ )
$$\mu _{1}
=\dot\imath\omega _{0},~~~ \mu _{2}=-\dot\imath \omega _0,$$
$$ \cos \varphi = \frac{\chi }{2} = \cos \omega _0 \tau +
\frac{\kappa }{\omega _0} \sin \Omega _0 \tau ,~~~ \Omega
=\omega_{0},$$
\noindent and from initial conditions (2) one has $~A_0=1$,
$~B_0=0$. The explicit expression for squeezing coefficient is
\begin{eqnarray}\label{eq.10}
K&=&U_{n-1}^2 +U_{n-2}^2 +\frac{2\kappa }{\omega _0}U_{n-1}^2
\sin 2\omega _0[t-(n-1)\tau ] - \chi U_{n-1} U_{n-2}
\nonumber\\
&+& \frac {4\kappa ^2}{\omega _0^2}U_{n-1}^2 (\sin \omega _0[t
-(n-1)\tau ])^2 -\frac {2\kappa }{\omega _0}U_{n-1}U_{n-2}
\sin{2\omega _0[t-(n-1/2)\tau ]}.
\end{eqnarray}
\noindent In the case of weak damping the squeezing coefficient
is determined by Eq. (9) with following parameters
\begin{eqnarray}\label{eq.11}
A_0&=&1-\dot\imath\gamma /2\Omega ,
\nonumber\\
 B_0&=&\frac{\dot\imath\gamma }{2\Omega },
\nonumber\\
 \Omega &=&(\omega ^2_0-\gamma ^2)^{1/2},
\nonumber\\
\frac{\chi }{2}&=&\cos\Omega \tau
+\frac{\kappa }{\Omega }\sin\Omega \tau ,
\nonumber\\
\mu _1&=&-\gamma +\dot\imath(\omega ^2_0-\gamma ^2)^{1/2},
\nonumber\\
\mu _2&=&-\gamma -\dot\imath(\omega ^2_0-\gamma ^2)^{1/2}.
\end{eqnarray}
\noindent One has the squeezing coefficient
\begin{eqnarray}\label{eq.12}
K &=&e^{-2\gamma t}\{K(\gamma =0)+
\frac{\gamma }{\Omega }
[\frac{2\kappa }{\Omega }U_{n-1}^2\cos 2\Omega \tau +
\frac{2\kappa ^2}{\Omega ^2}U_{n-1}^2\sin 2\Omega \tau -
\frac{2\kappa }{\Omega }U_{n-1}U_{n-2}\cos \Omega \tau
\nonumber \\
&+&(1-\frac{\kappa ^2}{\Omega ^2})U_{n-1}^2\sin 2\Omega (t
-\tau (n-2))+ U_{n-2}^2\sin 2\Omega (t-\tau (n-1))
\nonumber \\
&-&2\frac{\kappa }{\Omega }U_{n-1}^2\cos 2\Omega (t-(n-2)\tau )-
2U_{n-1}U_{n-2}\sin 2\Omega (t-(n-3/2)\tau )
\nonumber \\
&+&\frac{2\kappa }{\Omega }U_{n-1}U_{n-2}\cos 2\Omega (t-(n
-3/2)\tau )+\frac{\kappa ^2}{\Omega ^2}U_{n-1}^2\sin 2\Omega (t
-n\tau )]
\nonumber \\
&+&\frac{\gamma ^2}{2\Omega ^2}[(1
+\frac{2\kappa ^2}{\Omega ^2})U_{n-1}^2 +
U_{n-2}^2 - \chi U_{n-1}U_{n-2}+
\frac{2\kappa }{\Omega }U_{n-1}^2(\sin 2\Omega \tau -
\frac{\kappa }{\Omega }\cos 2\Omega \tau )
\nonumber \\
&-&2\frac{\kappa }{\Omega }U_{n-1}U_{n-2}\sin \Omega \tau +
\frac{2\kappa }{\Omega }U_{n-1}^2\sin 2\Omega (t-(n-1)\tau )
\nonumber \\
&-&\frac{2\kappa ^2}{\Omega ^2}U_{n-1}^2\cos 2\Omega (t-(n
-1)\tau )-\frac{2\kappa }{\Omega }U_{n-1}U_{n-2}\sin 2\Omega (t
-(n-1/2)\tau )
\nonumber \\
&+&\frac{\kappa ^2}{\Omega ^2}U_{n-1}^2\cos 2\Omega (t-n\tau )-
(1-\frac{\kappa ^2}{\Omega ^2})U_{n-1}^2\cos 2\Omega (t-(n-2)\tau )
\nonumber \\
&-&\frac{2\kappa }{\Omega }U_{n-1}^2\sin 2\Omega (t-(n-2)\tau )-
U_{n-2}^2\cos 2\Omega (t-(n-1)\tau )
\nonumber \\
&+&2U_{n-1}U_{n-2}\cos 2\Omega (t-(n-3/2)\tau )+
\frac{2\kappa }{\Omega }U_{n-1}U_{n-2}\sin 2\Omega (t-(n
-3/2)\tau ))]\}.
\end{eqnarray}
\noindent The squeezing phenomenon appears when the squeezing
coefficient starts to be less then 1. The force of delta--kicks
$~\kappa $ plays the main role in appearing of the squeezing phenomenon
at initial moments of time as can be seen from the previous formula,
with time increasing the damping begins to play the main role through
the exponential function. Let us mention for simplicity the
expression for squeezing coefficient in the case of one delta--kick of
frequency at the moment of time $~t=0$
$$K= e^{-2\gamma t}[K(\gamma =0)+ \frac{\gamma }{\Omega }(\sin 2\Omega t
+ \frac{4}{\Omega }(\kappa +{\gamma \over 4})\sin ^2\Omega t].$$
\noindent In the case of strong damping one has the following
expressions for the parameters
\begin{eqnarray}\label{eq.13}
 A_0&=&1/2+\dot\imath/2 +\gamma /2\Omega ,
\nonumber\\
B_0&=&1/2-\dot\imath/2-\gamma /2\Omega ,
\nonumber\\
\Omega &=&(\gamma ^2-\omega _0^2)^{1/2},
\nonumber\\
\mu _1&=&-\gamma +(\gamma ^2-\omega ^2_0)^{1/2},
\nonumber\\
\mu _2&=&-\gamma -(\gamma ^2-\omega ^2_0)^{1/2},
\nonumber\\
\frac{\chi }{2}&=&\cosh \Omega \tau
+\frac{\kappa }{\Omega }\sinh \Omega \tau .
\end{eqnarray}
\noindent Thus we have considered the parametric excitation of damped
oscillator in the frame of Caldirola--Kanai model and discussed the
influence of different regimes of damping on the squeezing coefficient
which describes squeezing phenomenon in the system. The parametric
excitation is chosen in the form of periodical $~\delta $--kicks of
frequency and the formulae for squeezing coefficient are expressed
through the Chebyshev polynomials.  It is necessary to add that different
aspects of the damped oscillator problem in the frame of Caldirola--Kanai
model were considered in Refs. \cite{[12]}--\cite{[17]}.  \\

\noindent  {\bf ACKNOWLEDGMENTS}\\

The research was supported by the Russian Science Foundation under grant
94-02-04715. The author wishes to thank the organizers of NATO Advanced Study
Institute "Electron Theory and Quantum Electrodynamics" for the support of the
participation in this conference.

\end{document}